\documentclass[
               twocolumn,
               noshowpacs,          
               nopreprintnumbers,     
               aps,                 
               prd,                 
               a4paper,             
               nofootinbib,         
               tightenlines,        
               floats,floatfix      
]{revtex4-1}
\usepackage{amsmath,amsfonts,amssymb,bbm}
\usepackage[mathscr]{eucal}
\usepackage{tensor}
\usepackage[T1]{fontenc}
\usepackage[utf8]{inputenc}
\usepackage{lmodern}
\usepackage{color}
\usepackage{collref}
\usepackage{hyperref}
\usepackage{graphicx}
\usepackage{float}
\usepackage{color}
\usepackage{tikz}
\usetikzlibrary{calc,decorations.markings}
\allowdisplaybreaks[1]
\hypersetup{
colorlinks=true,
linktoc=page,    
linkcolor=blue,
citecolor=blue,
urlcolor=blue,
colorlinks=true,
}

\usepackage{braket}
\usepackage{subfigure}
\usepackage{simplewick}
\usepackage{mathrsfs, extarrows}

\newcommand\td{\text{d}}
\newcommand{\p}{\partial}

\def\>{\rangle} \def\<{\langle}

\newcommand*\xbar[1]{%
  \hbox{%
    \vbox{%
      \hrule height 0.5pt 
      \kern0.3ex
      \hbox{%
        \kern-0.0em
        \ensuremath{#1}%
        \kern-0.0em
      }%
    }%
  }%
}
\def\bra#1{\left\langle #1\right|}
\def\ket#1{\left| #1\right\rangle}
\def\>{\rangle} 
\def\<{\langle} 

\begin{document}

\title{Quantum field theory in flat spacetime with multiple time directions}

\author{Bin Chen$^{1,2,3}$}
\email{chenbin1@nbu.edu.cn}

\author{Zezhou Hu$^2$}
\email{z.z.hu@pku.edu.cn}

\author{Xin-Cheng Mao$^2$}
\email{maoxc1120@stu.pku.edu.cn}

\affiliation{
Institute of Fundamental Physics and Quantum Technology, \\\&  School of Physical Science and Technology, \\ Ningbo University, Ningbo, Zhejiang 315211, China\\
\\
$^2$School of Physics, Peking University, \\No.5 Yiheyuan Rd, Beijing 100871, P.~R.~China\\
\\
$^3$Center for High Energy Physics, Peking University, \\No.5 Yiheyuan Rd, Beijing 100871, P.~R.~China\\}


\date{\today}

\begin{abstract}
This work serves as the generalization of the work \cite{Chen:2025acl}, where we investigated the quantum field theory in Klein space which has two time directions. We extend studies to the general spacetime $\mathbb{R}^{n,d-n}\,(n,d-n\geq2)$ in a similar manner; the ``length of time'' $q$ is regarded as the evolution direction of the system and additional modes beyond the plane wave modes are introduced in the canonical quantization. We show that this novel formulation is consistent with the analytical continuation of the results in Minkowski spacetime.
\end{abstract}

\maketitle
\section{Introduction}

In textbooks, a $d$-dimensional relativistic quantum field theory (QFT) is formulated in Minkowski spacetime $\mathbb{R}^{1,d-1}$, calculating the scattering amplitudes $\mathcal{A}_m(p_1,p_2,\cdots,p_m)$ among ingoing and outgoing particles. During canonical quantization, the field $\phi(t,\vec x)$ evolves over time $t$ and the Feynman correlation functions $C(x_1,\cdots,x_m)=\<0|T\phi(x_1)\cdots\phi(x_m)|0\>$ are calculated to derive the scattering amplitudes via the LSZ reduction formula,
\begin{equation}
    \mathcal{A}_m= \left[\prod_{k=1}^m i \int \td^4 x_k e^{-i p_k \cdot x_k}(-\p^2_k+M^2)\right]C(x_1,\cdots,x_m).
\end{equation}
When considering on-shell amplitudes ($p_k^2+M^2=0$), the outgoing particles are denoted as the particles with $p_k^0>0$ and the ingoing ones with $p_k^0<0$.

People used to consider the analytic continuation of these results from Minkowski spacetime to the spacetime $\mathbb{R}^{n,d-n}$ with general signature. While the Feynman correlation functions and the off-shell amplitudes remain sensible under the analytic continuation, the on-shell scattering amplitudes present very different properties. For instance, the on-shell three-point amplitudes of massless particles do not exist in Euclidean space ($n=0$) due to the nonexistence of on-shell states, and they universally vanish in Minkowski spacetime ($n=1$) for kinematic reasons, but they are nontrivial under the general signature when $n,d-n\geq2$ \cite{Britto:2005fq, Cachazo:2004kj, Arkani-Hamed:2008bsc, Benincasa:2007xk, Arkani-Hamed:2012zlh}.

Although the Feynman correlation functions and scattering amplitudes under the general signature can be obtained from analytic continuation, the intrinsic construction of a QFT in the space $\mathbb{R}^{n,d-n}\,(n,d-n\geq2)$ is lacking in the literature\footnote{In \cite{Melton:2024pre}, the authors constructed the $S$-vector in Klein space $\mathbb R^{2,2}$ as a Poincar\'e invariant vacuum state $|\mathcal{C}\>$ in the Hilbert space built on $\mathcal{J}$, where the Hilbert space $\mathcal{H}_\mathcal{J}=\mathcal{H}_{\mathcal{J}_+}\otimes\mathcal{H}_{\mathcal{J}_-}$. And the Hilbert spaces $\mathcal{H}_{\mathcal{J}_\pm}$ are built on AdS$_3/\mathbb{Z}$ slices, which is quite different from our starting point. More importantly, they did not explicitly show that their constructions lead to the results obtained from the analytical continuation from four-dimensional Minkowski spacetime.}. The main difficulty lies in the fact that such space has only one conformal boundary in contrast to Minkowski spacetime's two boundaries, which can be seen in the Penrose diagram Fig.\ref{fig:KleinPen}. Thereafter, the classical solution of the field composed of plane wave modes $e^{-i p \cdot x}$ is not enough for canonical quantization.

In this work, we intrinsically define a QFT in the space $\mathbb{R}^{n,d-n}\,(n,d-n\geq2)$ with additional modes beyond plane waves introduced to realize the canonical quantization, and we deduce novel annihilation conditions for two different vacuum states to calculate the Feynman correlation functions. We will focus on the scalar field theory in this work.

This work can be regarded as a different but equivalent formulation of QFT since all the results from our novel constructions match those in the textbook by analytic continuation. Moreover, it is the foundation for exploring many important questions, including scattering amplitudes \cite{Penrose:1967wn, Penrose:1968me, Penrose:1985bww, Parke:1986gb, Dunajski:2001ea, Witten:2003nn, Arkani-Hamed:2009hub, Monteiro:2020plf}, path integral\cite{Schwinger:1958mma, Heckman:2022peq}, self-dual gravity \cite{Penrose:1976jq, KO198151, flathspaces, Eguchi:1978xp}, black holes \cite{Crawley:2021auj}, the Unruh effect \cite{Santos:2023pwg}, etc. 

Flat space with multiple time directions $\mathbb R^{n, d-n}$ also plays a unique role in studying holography \cite{tHooft:1993dmi, Susskind:1994vu}. On the one hand, the $n=2$ case can be related to the celebrated AdS/CFT \cite{Maldacena:1997re, Gubser:1998bc, Witten:1998qj} by viewing AdS$_{d-1}$ as hyperbolic slices embedded in the space $\mathbb{R}^{2,d-2}$ \cite{Ball:2019atb, Casali:2022fro, Iacobacci:2022yjo, Melton:2023dee, Bu:2023cef}. On the other hand, since the flat space $\mathbb R^{n, d-n}$ with $n, d-n \geq 2$ has only one conformal boundary, it is more suitable and convenient for investigations of flat holography \cite{Susskind:1998vk, Polchinski:1999ry,deBoer:2003vf, Arcioni:2003xx, Arcioni:2003td, Solodukhin:2004gs, Barnich:2006av, Guica:2008mu, Barnich:2009se, Barnich:2010eb, Bagchi:2010zz, Bagchi:2012xr} from both the Carrollian approach \cite{Bagchi:2012cy, Bagchi:2022nvj, Banerjee:2022ime, Bagchi:2023fbj, Alday:2024yyj, Donnay:2022aba, Donnay:2022wvx, Chen:2023naw, Chen:2023pqf, Bagchi:2023cen, Bagchi:2022emh, Bagchi:2024unl} and the Celestial approach \cite{Pasterski:2016qvg, Pasterski:2017kqt, Pasterski:2017ylz, Raclariu:2021zjz, Pasterski:2021rjz, Pasterski:2021raf, Strominger:2017zoo, Atanasov:2021oyu, Melton:2023hiq, Melton:2023bjw, Melton:2024jyq, Melton:2024pre, Duary:2024cqb, Bhattacharjee:2021mdc}. 

\section{The space $\mathbb{R}^{n,d-n}\,(n,d-n\geq2)$} \label{sec:RevKlein}

\begin{figure}[htbp]
    \centering
    \includegraphics[width=0.5\linewidth]{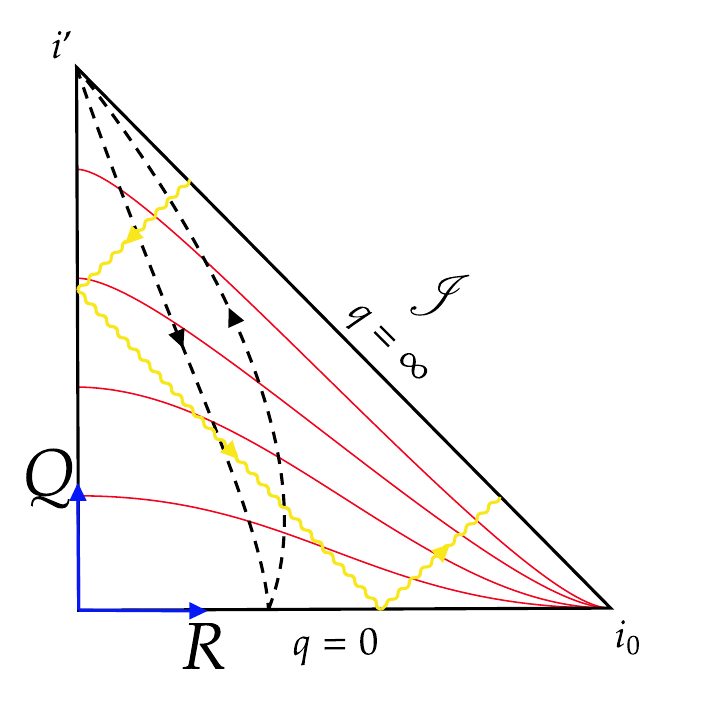}
    \caption{The Penrose diagram of flat space $\mathbb{R}^{n,d-n}\,(n,d-n\geq2)$. The infinite coordinate ranges $q, r \in [0, \infty)$ are conformally mapped to finite ranges $Q, R \in [0, \pi)$. The conformal boundary emerges at the singularity of $\Omega^{-1}$, with null infinity $\mathscr{I}$ located at $Q + R = \pi$. The endpoints $(Q, R) = (0, \pi)$ and $(Q, R) = (\pi, 0)$ correspond to the spacelike infinity $i_0$ and timelike infinity $i'$, respectively.}
    \label{fig:KleinPen}
\end{figure}

The space $\mathbb{R}^{n,d-n}\,(n,d-n\geq2)$ has a metric of the form 
\begin{equation}
    \begin{aligned}
        \td s^2 &= - \sum_{i=0}^{n-1}(\td x^i)^2 +\sum_{j=n}^{d-1} (\td x^j)^2\\
        &= - \td q^2 -q^2 \td \vec \psi^2+\td r^2+r^2\td\vec\varphi^2\\
        &=\Omega^{-2} (- \td Q^2  - \sin^2 Q \td \vec\psi^2 + \td R^2+\sin^2 R\td \vec\varphi^2),
    \end{aligned}
\end{equation}
with $\vec \psi,\,\vec \varphi$ being $(n-1)$-dimensional timelike and $(d-n-1)$-dimensional spacelike spherical coordinates.
Its Penrose diagram, shown in Fig.\ref{fig:KleinPen}, with conformal factor
\begin{equation}
    \Omega = 2 \left| \cos \left( \frac{R + Q}{2} \right) \cos \left( \frac{R - Q}{2} \right) \right|\,,
\end{equation}
can be obtained by
\begin{equation}
    \begin{split}
        Q  &= \arctan (q+r) + \arctan (q-r)\,, \\
        R &= \arctan (q+r) - \arctan (q-r)\,.
    \end{split}
\end{equation}

In the Penrose diagram, the black dashed line and the yellow wavy line (45°) denote the worldlines of massive and massless particles, respectively. Unlike the Minkowski space $\mathbb R^{1,d-1}$, which has two null boundaries, the space $\mathbb{R}^{n,d-n}\,(n,d-n\geq2)$ possesses only a single null infinity. This structural difference necessitates replacing the conventional $S$-matrix with an $S$-vector to describe scattering processes, following the discussion in \cite{Witten:2001kn}.

Then in our construction, the fields evolve over the coordinate $q$, the ``length'' of the time, from the original point to the conformal boundary.
The associated coordinate system is equipped with a metric
\begin{equation} \label{eq:polar}
    \td s^2 = - \td q^2+\td s^2_{\Sigma_q}\,,
\end{equation}
where the induced metric of $q = $ constant hypersurface $\Sigma_q$ is given by 
\begin{equation}
    \td s^2_{\Sigma_q} = - q^2 \td \vec\psi^2 + \td \vec x^2, \quad \vec x = (x^{n+1}, x^{n+2},\cdots, x^d).
\end{equation}

The momentum space inherits the same geometric structure as its dual space $\mathbb R^{n,d-n}$. Consequently, we can parametrize the $d$-momentum $p^\mu=(\vec p^{(n)}, \vec{p})$ using analogous coordinates
\begin{align}
        \vec p^{(n)} & = (\omega \cos\psi_1,\omega \sin\psi_1\cos\psi_2,\cdots,\omega\sin\psi_1\cdots\sin\psi_{n-1}), \notag \\
        \vec{p}&=(p^{n+1}, p^{n+2},\cdots,p^{d-n-1})
\end{align}
with $\omega>0$. Then the $SO (n, d-n)$-invariant momentum measure can be defined in a similar manner to that in Minkowski space
\begin{align}
    \int \td^d p \delta (p^2 + M^2) & = \int \omega^{n-1} \td \omega \td^{n-1} \vec\psi_p \td^{d-n} \vec p \delta (- \omega^2 + \omega_{\vec{p}}^2) \notag \\
    & = \frac12 \int \td^{n-1}\vec \psi_p \int \td^{d-n} \vec p\, \omega_{\vec p}^{n-2}
\end{align}
where $\omega_{\vec p}= \sqrt{|\vec p|^2 + M^2}$ is the on-shell energy.

\section{Scalar field theory} \label{sec:CanonicalQ}

A real scalar field is described by the Lagrangian
\begin{equation}
        \mathcal{L}  = - \frac12 (\p_{\mu} \phi \p^{\mu} \phi + M^2 \phi^2) - V (\phi)\,,
\end{equation}
where $V (\phi)$ denotes the small interacting term, which can be ignored at this stage.
By using the coordinates \eqref{eq:polar}, the free field can be expanded as
\begin{equation}
    \begin{split}
        \phi = & \sum_{l = 0}^\infty \sum_{\{\mu\}} Y^{(n)}_{l,\{\mu\}} (\vec \psi) \int \frac{\td^{d-n} \vec p}{(2 \pi)^{d-n}} e^{i \vec p \cdot \vec x} \\
        & \times \left[ a^{(J)}_{\vec p; l, \{\mu\}} j^{(n)}_l (\omega_{\vec p} q) + a^{(N)}_{\vec p; l, \{\mu\}} n^{(n)}_l (\omega_{\vec p} q) \right].
    \end{split}
\end{equation}
where $Y^{(n)}_{l, \{\mu\}}$ denotes the hyperspherical harmonic \cite{wen1985some}, and the functions $f (u) = j_l^{(n)} (u)$ and $n_l^{(n)} (u)$ with $u=\omega_{\vec p} q$ obeying the equation
\begin{equation} \label{eq:sBessel(n)}
    u^2 f'' (u) + (n-1) u f'(u) + [u^2 - l (l + n -2) ] f (u)=0\,,
\end{equation}
are given by
\begin{equation}
    j_l^{(n)} (u) = \frac{J_{l + \frac{n-2}{2}} (u)}{(u/2)^{\frac{n-2}{2}}}\,, \qquad n^{(n)}_l (u) =  \frac{N_{l + \frac{n-2}{2}} (u)}{(u/2)^{\frac{n-2}{2}}}\,,
\end{equation}
where $J_\nu(u)$ and $N_\nu(u)$ are the first- and second-kind Bessel functions, respectively.
In particular, the $n = 2$ case reduces to conventional Bessel functions. In the $n=3$ case, the solution $f(u)$ is the spherical ($S^2$) Bessel function. The functions $j_l^{(n)}, n^{(n)}_l$ for larger $n$ are the generalization of the Bessel function, for simplicity, we call them ``$S^{n-1}$ Bessel functions''.

The conjugate momentum is defined as
\begin{equation}
    \pi_q = i^{n-1} q^{n-1} \sqrt{h^{(S)}_{n-1}} \p_q \phi
\end{equation}
where $h^{(S)}_{n-1}$ denotes the determinant of the metric of a unit sphere $S^{n-1}$. At $q = $ constant hypersurfaces, the canonical quantization condition is
\begin{equation}
    \begin{split}
        [\phi (q, \vec \psi_2, \vec x_2), \pi_q (q, \vec \psi_1, \vec x_1)] & = i^{n-1} \delta (\vec \psi_{12}) \delta (\vec x_{12}), \\
        [\phi (q, \vec \psi_1, \vec x_1), \phi (q, \vec \psi_2, \vec x_2)] & = 0,
    \end{split}
\end{equation}
becomes
\begin{equation}\label{eq:cancommu(n)}
    \left[ a^{(J)}_{\vec p; l, \{\mu\}}, a^{(N)}_{\vec p'; l', \{\mu'\}} \right] = - \frac{\pi}{2} (\omega_{\vec p}/2)^{n-2} \delta_{l, l'} \delta_{\{\mu\}, \{\mu'\}} \delta (\vec p + \vec p')\,,
\end{equation}
where we used the identity for the $S^{n-1}$ Bessel functions
\begin{equation}
    j^{(n)}_l (\omega_{\vec p} q) (\omega_{\vec p} q) \overset{\longleftrightarrow}{(q^{n-1} \p_q)} n^{(n)}_l (\omega_{\vec p} q) = \frac{2}{\pi} (\omega_{\vec p}/2)^{2-n}.
\end{equation}

\section{Vacuum states and correlation functions}

Since our system evolves from $q=0$ to $q=\infty$, we interpret the Feynman correlation functions as $q$-ordered operator vacuum expectations similar to those in Minkowski spacetime, 
\begin{equation}
    C^{(n)} (x_1,\cdots,x_m)=\<\infty|\mathcal{Q}\phi(x_1)\cdots\phi(x_m)|0\>\,.
\end{equation}

The state at the original point $|0\>$ should be constrained by the regularity condition
\begin{equation} \label{eq:regquant}
    \phi (q, \psi, \vec x) \ket{0} \Big|_{q = 0} = \text{finite}\,.
\end{equation}
Since the Neumann function modes $n_l^{(n)}(u)$ is divergent at $q=0$, the regularity condition becomes
\begin{equation} \label{eq:regular}
    a^{(N)}_{n, \vec p} \ket{0} = 0 \quad (\forall n, \vec p)\,.
\end{equation}
This is analogous to the definition of asymptotic states in conformal field theory \cite{DiFrancesco:1997nk}.
Since the state $|0\>$ is annihilated by the Neumann function coefficient, we will call it the Neumann vacuum in contrast with the Hankel vacuum $\<\infty|$, which will be defined in (\ref{eq:InftyStateAnnihilation}) later.

As the vector $\partial_q$ is not a (conformal) Killing vector, the vacuum state $\<\infty|$ at $q\rightarrow\infty$ is not related to the Neumann vacuum $|0\>$ at all. The field $\phi$ at the asymptotic infinity behaves as
\begin{multline*}
    \phi  = \sum_{l = 0}^\infty \sum_{\{\mu\}} Y^{(n)}_{l,\{\mu\}} (\vec \psi^{(n-1)}) \int \frac{\td^{d-n} \vec p}{(2 \pi)^{d-n}} e^{i \vec p \cdot \vec x} \\
    \times \left[ a^{(H^{(1)})}_{\vec p; l, \{\mu\}} h^{(1), (n)}_l (\omega_{\vec p} q)+ a^{(H^{(2)})}_{\vec p; l, \{\mu\}} h^{(2), (n)}_l (\omega_{\vec p} q) \right] \\
    \underset{q \to \infty}{\longrightarrow} \sum_{l, \{\mu\}} Y^{(n)}_{l,\{\mu\}} (\vec \psi^{(n-1)}) \int \frac{\td^{d-n} \vec p}{(2 \pi)^{d-n}} \frac{e^{i \vec p \cdot \vec x}}{\sqrt{\pi (\omega_{\vec p}q)^{n-1}}} \\
    \times \Big[ a^{(H^{(1)})}_{\vec p; l, \{\mu\}} e^{i\left(\omega_{\vec p}q-\frac{\pi}{2} (l - 1 + \frac n2) -\frac{\pi}{4}\right)} + a^{(H^{(2)})}_{\vec p; l, \{\mu\}} e^{-i \left(\omega_{\vec p}q-\frac{\pi}{2} (l - 1 + \frac n2) -\frac{\pi}{4}\right)} \Big]
\end{multline*}
where
\begin{equation}
    a^{(H^{(1)})}_{\vec p; l, \{\mu\}} =\frac{a^{(J)}_{\vec p; l, \{\mu\}} -i a^{(N)}_{\vec p; l, \{\mu\}}}{2}\,,\quad a^{(H^{(2)})}_{\vec p; l, \{\mu\}} =\frac{a^{(J)}_{\vec p; l, \{\mu\}} +i a^{(N)}_{\vec p; l, \{\mu\}}}{2}\,\,,
\end{equation}
and the Hankel function modes are
\begin{equation}
    h_l^{(1),(n)}(u) = j_l^{(n)}(u) + i n_l^{(n)}(u), \, h_l^{(2),(n)}(u) = j_l^{(n)}(u) - i n_l^{(n)}(u)\,.
\end{equation}
The modes $e^{i\omega_{\vec p}q}$ and $e^{-i\omega_{\vec p}q}$ are similar to $e^{i\omega t}$ and $e^{-i\omega t}$ in the Minkowski case. Thus, the vacuum state $\<\infty|$ should satisfy
\begin{equation}\label{eq:InftyStateAnnihilation}
    \bra{\infty} a^{(H^{(1)})}_{\vec p; l, \{\mu\}} = 0\,, 
\end{equation}
and we will call it the Hankel vacuum since the state $\<\infty|$ is annihilated by the Hankel function coefficient acting from the right side.

Now we can calculate the Feynman correlation functions via the Wick contraction, based on the commutation relations \eqref{eq:cancommu(n)}, the annihilation conditions of Neumann vacuum (\ref{eq:regular}) and the Hankel vacuum (\ref{eq:InftyStateAnnihilation}). Let us consider the two-point propagator $\Delta (x - x') =\langle \infty| \mathcal{Q} \phi(q,\psi,\vec x)\phi(q',\psi',\vec x')|0\rangle$. Set $q>q'$ without loss of generality, it is
\begin{equation} \label{eq:2ptnm}
    \begin{split}
        &\bra{\infty} \phi (q, \vec \psi, \vec x) \phi (q', \vec \psi', \vec x') \ket{0} / \bra{\infty} 0 \> \\
        = &\frac{i \pi^{1- \frac n2}}{8} \int \frac{\td^{d-n} \vec p}{(2 \pi)^{d-n}} e^{i \vec p\cdot(\vec x - \vec x')} (\omega_{\vec p} /2)^{n-2} h^{(2), (n)}_0 (\omega_{\vec p} \Delta q)\,,
    \end{split}
\end{equation}
with
\begin{equation}
    \Delta q=\sqrt{q^2+q'^2-2q q'(\vec\psi\cdot\vec\psi')}
\end{equation}

The two-point Feynman correlation function from analytical continuation is
\begin{multline} \label{eq:analyticn}
    \frac{\bra{\infty} \phi (q, \vec \psi, \vec x) \phi (q', \vec \psi', \vec x') \ket{0}} {\bra{\infty} 0 \>}  \propto \int \frac{\td^{d} p}{(2\pi)^{d}} \frac{e^{i p \cdot (x - x')}}{p^2 + M^2 - i \epsilon} \\
    = \frac{(n - 2)!! A(S^{n-1})}{\mathcal{N}_n} \int \frac{\td^{d-n} \vec p}{(2\pi)^d} e^{i \vec p \cdot (\vec x - \vec x')} \mathcal{I}^{(n)}_{\vec p}, 
\end{multline}
where the coefficient $\mathcal{N}_n = 2^{\frac n2}$ or $2^{\frac{n+1}{2}} / \Gamma (\frac12)$ when $n$ is even or odd, respectively.
The integral $\mathcal{I}^{(n)}_{\vec p}$ is calculated as
\begin{equation} \label{eq:In}
    \begin{split}
        \mathcal{I}^{(n)}_{\vec p} & = \int_0^\infty \td \omega \omega^{n-1} \frac{j^{(n)}_0 (\omega \Delta q)}{- \omega^2 + \omega_{\vec p}^2 - i \epsilon} \\
        & = \frac{\pi i}{2} \omega_{\vec p}^{n-2} h^{(2), (n)}_0 (\omega_{\vec p} \Delta q),
    \end{split}
\end{equation} 
which means the result from our construction \eqref{eq:2ptnm} is consistent with that from analytical continuation \eqref{eq:analyticn}. 

The details of computing the correlation functions are presented in Appendix \ref{ap:CorrelationFunc}.

\section{Conclusions and discussions}\label{sec:ConDiss}

In this work, we introduced how to canonical quantize a field theory in a generic flat spacetime $\mathbb{R}^{n,d-n} (n,d-n\geq2)$ with multiple time directions. We selected the ``length of time'' $q$ as the evolution parameter such that the field can be expanded in terms of the Bessel functions and the Neumann functions. Though the Neumann functions are divergent at the original point, they are necessary in canonical quantization. The calculations in this work resemble the previous discussions in Klein space for $n=2$ \cite{Chen:2025acl}, which also has only one asymptotic boundary. We imposed the regularity condition to constrain the Neumann vacuum state $|0\>$. On the other hand, we also define the Hankel vacuum state $\<\infty|$ which is associated with the asymptotic region $q\rightarrow\infty$, where the mode expansions behave as $e^{\pm i \omega_{\vec p}q}$. By using these constructions, we explicitly calculated the free two-point function \eqref{eq:2ptnm}, which is consistent with the covariant one \eqref{eq:analyticn}. The agreement between the results from our intrinsic construction and those from analytic continuation justifies the correctness of our formalism.

The novel modes associated with the Neumann functions need further study. They may play an essential role in flat holography. One unanswered question in flat holography concerns the extrapolation dictionary between the operators in boundary field theory and the bulk fields. It would be interesting to investigate the role of these novel modes in the bulk reconstruction in flat holography.

As discussed in \cite{Chen:2025acl}, the LSZ reduction formula, path-integral quantization, and perturbative calculation of interacting theory are all realizable in the space $\mathbb{R}^{n,d-n}$. Due to the limitation of this manuscript, we only demonstrate the LSZ reduction formula in Appendix \ref{ap:LSZ}. We also give some details on the calculations of the correlation functions in Appendix \ref{ap:CorrelationFunc}.

\acknowledgments
We would like to thank Yu-fan Zheng for inspiring discussions and thank Yu-ting Wen, Jie Xu, and Zhi-jun Yin for the valuable suggestions on the manuscript. This research is supported in part by NSFC Grants No. 11735001 and No. 12275004. 

\appendix

\section{The interacting real scalar field and LSZ reduction formula}\label{ap:LSZ}

For the interacting real scalar, the potential $V (\phi)$ in the Lagrangian is nonvanishing, such that the equation of motion should be modified to include potential terms. Here we adopt the Heisenberg picture in which the mode expansion can be formally performed in a similar form to that of free field theory
\begin{multline}
    \phi = \sum_{l = 0}^\infty \sum_{\{\mu\}} Y^{(n)}_{l,\{\mu\}} (\vec \psi) \int \frac{\td^{d-n} \vec p}{(2 \pi)^{d-n}} e^{i \vec p \cdot \vec x} \\
    \times \left[ a^{(J)}_{\vec p; l, \{\mu\}} (q) j^{(n)}_l (\omega_{\vec p} q) + a^{(N)}_{\vec p; l, \{\mu\}} (q) n^{(n)}_l (\omega_{\vec p} q) \right]
\end{multline}
But now the coefficients $a^{(J)}(q), a^{(N)}(q)$ are no longer independent of the coordinate $q$. 

At the quantum level, $\phi(q\rightarrow0)|0\>$ still needs to be finite, leading to
\begin{equation} \label{eq:regularint}
    a^{(N)}_{\vec \psi_p, \vec p} (q = 0) \ket{0} = 0 \qquad \forall \vec \psi_p, \vec p\,.
\end{equation}
with 
\begin{equation}
    a^{(N)}_{\vec \psi_p, \vec p} (q) = \frac{\omega_{\vec p}^{2-n} \mathcal{N}_n}{2^{-n/2} \pi} \sum_{l = 0}^\infty \sum_{\{\mu\}} i^l Y^{(n)}_{l, \{\mu\}} (\vec \psi_p) a^{(N)}_{\vec p; l, \{\mu\}} (q).
\end{equation}
Here, the coefficient $\mathcal{N}_n$ is the same as that in Eq.\eqref{eq:analyticn}.

In general, $a^{(N)}_{\vec \psi_p, \vec p} (q = \infty) \ket{0} \neq 0$ as $a^{(N)}_{\vec \psi_p, \vec p} (\infty)$ differs from $a_{\vec \psi_p, \vec p}^{(N)} (0)$ in a nontrivial way. Their difference can be computed as
\begin{multline} \label{eq:diffaNinfaN0}
    a^{(N)}_{\vec \psi_p, \vec p} (\infty)-a^{(N)}_{\vec \psi_p, \vec p} (0) =\<e^{-i p x}, \phi(x) \> \Big|_{q = 0}^{\infty}\\
    = \int \td^d x e^{-i p \cdot x} (- \p^2 + M^2) \phi (x) \,.
\end{multline}
Here, we used the expansion of $n$-dimensional plane wave
\cite{wen1985some}
\begin{multline}
    e^{i \vec p^{(n)} \cdot \vec x^{(n)}} 
    = 2^{- \frac{n}{2} -1} \mathcal{N} _n \Gamma \left( \frac{n}{2} - 1 \right) \sum_{l = 0}^\infty i^l (n+2l-2) \\ 
    \times j_l^{(n)} (|\vec p^{(n)}| |\vec x^{(n)}|) C_l^{(\frac{n-2}{2})} \left( \frac{\vec p^{(n)} \cdot \vec x^{(n)}}{|\vec p^{(n)} | | \vec x^{(n)}|} \right)
\end{multline}
where $C_l^{(\frac{n-2}{2})}$ is the Gegenbauer polynomial. Specifically, $C_l^{\frac 12}(\vec \psi \cdot \vec \psi')$ (namely $n=3$) is the Legendre polynomial $P_l(\vec \psi \cdot \vec \psi')$. In the derivation of \eqref{eq:diffaNinfaN0}, we also used the expansion formula for the Gegenbauer polynomial \cite{avery2002hyperspherical}
\begin{equation} \label{eq:C=YY}
    C_l^{(\frac{n-2}{2})} (\vec \psi \cdot \vec \psi') = \frac{(n-2) A(S^{n-1})}{2l + n-2} \sum_{\{ \mu\}} Y^{(n)*}_{l, \{\mu\}} (\vec \psi) Y^{(n)}_{l, \{\mu\}} (\vec \psi')\,
\end{equation}
with $A(S^{n-1}) = \frac{2 \pi^{\frac n2}}{\Gamma (\frac n2)}$ being the area of $S^{n-1}$, the definition of the Klein-Gordon inner product
\begin{equation}
    \< \phi^1, \phi^2 \> = \int_{\Sigma_q} \td^{n-1} \vec \psi \td^{d-n} \vec x \phi^1 \overset{\longleftrightarrow}{(q^{n-1} \p_q)} \phi^2,
\end{equation}
and the orthonormal condition for the hyperspherical harmonics \cite{avery2002hyperspherical}
\begin{equation}
    \int \td^{n-1} \vec \psi Y^{(n)*}_{l, \{\mu\}} (\vec \psi) Y^{(n)}_{l', \{\mu'\}} (\vec \psi) = \delta_{l, l'} \delta_{\{\mu\}, \{\mu'\}}.
\end{equation}

The scattering amplitude is defined by the inner product between the scattering state living on the conformal boundary and the $S$-vector $|S\rangle\equiv |0\rangle$ (originally mentioned in section \ref{sec:RevKlein}), or more explicitly
\begin{equation}
    \mathcal{A}_m := \langle p_1, \cdots ,p_m|S\rangle = \langle \infty|a_{p_1}(\infty) \cdots a_{p_m}(\infty)|0\rangle,
\end{equation}
where the scattering state defined at the asymptotic region $q \rightarrow \infty$ is given by
\begin{equation}\label{eq:AsymScattStat}
    \begin{aligned}
        \langle p_1, \cdots, p_m|&=\langle \infty|a_{p_1}(\infty) \cdots a_{p_m}(\infty).
    \end{aligned}
\end{equation}
The explicit expression for the counterpart of $a_p$ in Minkowski space is the Klein-Gordon inner product between the plane wave and the field $\phi (x)$. In the space $\mathbb{R}^{n,d-n}\,(n,d-n\geq2)$, $a_p$ can be similarly defined as the plane wave generator $\<e^{-i p x}, \phi (x)\>$ in the asymptotic regions $q \rightarrow 0, \infty$,
\begin{equation} \label{eq:aInScattering}
    \left[ a_{p}(q) := \< e^{- i p  x}, \phi (x) \> = a^{(N)}_{\vec \psi_p, \vec p} (q) \right] \Big|_{q \rightarrow 0, \infty}.
\end{equation}
Therefore, the amplitude becomes
\begin{equation} \label{eq:LSZred}
    \begin{aligned}
        \mathcal{A}_m&=\<\infty|\mathcal{Q} \prod^m_{k = 1} \left(a^{(N)}_{\vec \psi p_k, \vec p_k} (\infty)-a^{(N)}_{\vec \psi p_k, \vec p_k} (0) \right) |0\> \\
        & =\langle \infty| \mathcal{Q} \prod^m_{k = 1} \< e^{-i p_k \cdot x_k}, \phi (x_k) \> \Big|_{q = 0}^{\infty} |0\rangle
    \end{aligned}
\end{equation}
where we use the regularity condition \eqref{eq:regularint} to add the $a^{(N)}_{\theta_i,\vec p_i}(0) \ket{0}$ term in the first equality. Finally, by using Eq.(\ref{eq:diffaNinfaN0}), we have the LSZ reduction formula.
\begin{equation}
        \mathcal{A}_m =  \left[ \prod^m_{k = 1} \int \td^d x_k e^{-i p_k \cdot x_k}(-\p_k^2+M^2) \right] C^{(n)} (x_1,\cdots,x_m) \\
\end{equation}

\section{Calculations of the correlation functions}\label{ap:CorrelationFunc}

In this appendix, we present the computation details of the correlation functions.

In the derivation of the \eqref{eq:2ptnm}, we have used the identity
\begin{multline}
    h^{(2), (n)}_0 (\omega_{\vec p} \Delta q) = \Gamma \left( \frac n2 -1 \right) \sum_{l = 0}^\infty (2l + n -2) \\
    \times h^{(2), (n)}_l (\omega_{\vec p} q) j^{(n)}_l (\omega_{\vec p} q') C_l^{(\frac{n-2}{2})} (\vec \psi \cdot \vec \psi')\,,
\end{multline}
and the Eq.\eqref{eq:C=YY}.

In the derivation of the \eqref{eq:analyticn}, we used the following identity \cite{avery2002hyperspherical}
\begin{equation}
    \begin{split}
        \int \td^{n-1} \Omega e^{-i \vec a \cdot \vec b} 
        & = \frac{(n - 2)!! A(S^{n-1})}{\mathcal{N}_n} j_0^{(n)} (|\vec a| |\vec b|)
    \end{split}
\end{equation}
for any $n$-dimensional vectors $\vec a, \vec b$.

To derive the Eq.\eqref{eq:In}, we will used the recursion relation
\begin{equation} \label{eq:recurbn}
    \mathfrak b_0^{(n)} (x) = 
    \begin{cases}
        & \left(- \frac{2}{x} \frac{\td}{\td x} \right)^{k -1} \mathfrak b_0^{(2)} (x), \quad n = 2k \\
        & \left(- \frac{2}{x} \frac{\td}{\td x} \right)^{k -1} \mathfrak b_0^{(3)} (x), \quad n = 2k+1
    \end{cases}
\end{equation}
for $S^{n-1}$ spherical Bessel functions $\mathfrak b_\nu^{(n)} = j_\nu^{(n)}, n_\nu^{(n)}, h_\nu^{(1), (n)}, h_\nu^{(2), (n)}$ to reexpress the integral $\mathcal{I}^{(n)}_{\vec p}$ by $\mathcal{I}^{(2)}_{\vec p}$ or $\mathcal{I}^{(3)}_{\vec p}$ for even or odd $n$
\begin{equation} \label{eq:Ipn}
    \mathcal{I}^{(n)}_{\vec p} =
    \begin{cases}
        & \left(\frac{-2}{\Delta q} \frac{\p}{\p (\Delta q)} \right)^{k -1} \mathcal{I}^{(2)}_{\vec p}, \quad n = 2k \\
        & \left(\frac{-2}{\Delta q} \frac{\p}{\p (\Delta q)} \right)^{k -1} \mathcal{I}^{(3)}_{\vec p}, \quad n = 2k+1
    \end{cases}
\end{equation}
The relation \eqref{eq:recurbn} can be derived from $\frac{\mathcal{B}_{\nu + 1} (x)}{x^\nu} = - \frac{\td}{\td x} \left[ \frac{\mathcal{B}_{\nu} (x)}{x^\nu} \right]$ for Bessel functions $\mathcal{B}_\nu = J_\nu, N_\nu, H^{(1)}_\nu, H^{(2)}_\nu$. 
Therefore, we only need to compute $\mathcal{I}^{(2)}_{\vec p}$ and $\mathcal{I}^{(3)}_{\vec p}$. 

Note that the $\mathcal{I}^{(2)}_{\vec p}$ is computed in Appendix B of \cite{Chen:2025acl}
\begin{equation}
    \mathcal{I}^{(2)}_{\vec p} = \frac{\pi i}{2} h^{(2), (2)}_0 (\omega_{\vec p} \Delta q)\,; 
\end{equation}
while the calculation for the $n=3$ case is relatively simple:
\begin{equation}
    \begin{split}
        \mathcal{I}^{(3)}_{\vec p} & = \int_0^\infty \td \omega \omega^2 \frac{j^{(3)}_0 (\omega \Delta q)}{- \omega^2 + \omega_{\vec p}^2 - i \epsilon} \\
        & = \int_{-\infty}^\infty \td \omega \omega (\Delta q)^{-1} \frac{i \pi^{-1/2} e^{i \omega \Delta q}}{\omega^2 - \omega_{\vec p}^2 + i \epsilon} \\
        & = \frac{\pi i}{2} \omega_{\vec p} h^{(2), (3)}_0 (\omega_{\vec p} \Delta q)
    \end{split}
\end{equation}
where we used the expression of zeroth-ordered spherical Bessel function $j^{(3)}_0 (x) = \frac{e^{ix} - e^{-ix}}{i \sqrt{\pi} x}, h^{(2), (n=3)}_0 (x) = \frac{2}{\sqrt{\pi} i} \frac{e^{-ix}}{x}$ and used the residue theorem. And thereafter, the Eq.\eqref{eq:In} is proved by using \eqref{eq:recurbn} and \eqref{eq:Ipn}.

\bibliography{ref}

\end{document}